\documentclass[prd,amsmath,amssymb,nofootinbib]{revtex4} 
\usepackage{graphicx} 
\usepackage{amsmath}
\usepackage{amsfonts,amsbsy}
\usepackage{amssymb}

\newcommand{\be}{\begin{equation}}
\newcommand{\ee}{\end{equation}}
\newcommand{\bea}{\begin{eqnarray}}
\newcommand{\eea}{\end{eqnarray}}
\newcommand{\tr}{\mathrm{tr}\,}
\renewcommand{\d}{\mathrm{d}}
\def\dd#1{\frac{\mathrm{d}^2#1}{(2\pi)^2}}

\def\lsim{\mathrel{\rlap{\lower4pt\hbox{\hskip1pt$\sim$}}
    \raise1pt\hbox{$<$}}}                
\def\gsim{\mathrel{\rlap{\lower4pt\hbox{\hskip1pt$\sim$}}
    \raise1pt\hbox{$>$}}}                

\begin{document}
\title{Fluctuations of the gluon distribution from the small-x
  effective action}

\author{Adrian Dumitru}
\affiliation{Department of Natural Sciences, Baruch College, CUNY,
17 Lexington Avenue, New York, NY 10010, USA}
\affiliation{The Graduate School and University Center, The City
  University of New York, 365 Fifth Avenue, New York, NY 10016, USA}

\author{Vladimir Skokov}
\affiliation{RIKEN/BNL Research Center, Brookhaven National
  Laboratory, Upton, NY 11973, USA}

\begin{abstract}
The computation of observables in high energy QCD involves an average
over stochastic semi-classical small-$x$ gluon fields. The weight of
various configurations is determined by the effective action. We
introduce a method to study fluctuations of observables, functionals
of the small-$x$ fields, which does not explicitly involve dipoles.
We integrate out those fluctuations of the semi-classical gluon field
under which a given observable is invariant. Thereby we obtain the
effective potential for that observable describing its fluctuations
about the average. We determine explicitly the effective potential for
the covariant gauge gluon distribution both for the
McLerran-Venugopalan (MV) model and for a (non-local) Gaussian
approximation for the small-$x$ effective action. This provides
insight into the correlation of fluctuations of the number of hard
gluons versus their typical transverse momentum. We find that the
spectral shape of the fluctuations of the gluon distribution is
fundamentally different in the MV model, where there is a pile-up of
gluons near the saturation scale, versus the solution of the small-$x$
JIMWLK renormalization group, which generates essentially scale
invariant fluctuations above the absorptive boundary set by the
saturation scale.
\end{abstract}

\maketitle

\section{Introduction}

High-energy scattering in QCD at fixed transverse momentum scales
probes strong color fields, i.e.\ the regime of high gluon
densities~\cite{GLR}. In the high-energy limit physical observables,
such as the forward scattering amplitude of a dipole from a hadron or
nucleus, are typically expressed in terms of expectation values of
various Wilson line operators $O$; see, for example,
Ref.~\cite{Weigert:2005us}. The expectation value $\langle O\rangle$
corresponds to a statistical average\footnote{Kovner describes this as
  an average over the Hilbert space of the target, i.e.\ that the
  weight $W[A^+]\equiv \exp(-S[A^+])$ which determines the probability
  for a given configuration of $A^+$ is analogous to the modulus
  squared of the wave function of the target~\cite{Kovner:2005pe}.}
over the distribution of ``small-x gluon fields''. For example, if
quantum corrections are neglected this distribution is commonly
described by the McLerran-Venugopalan (MV) model~\cite{MV}:
\bea
-\nabla_\perp^2 A^+(x^-,x_\perp) &=& g\rho(x^-,x_\perp)~, \\
Z = \int {\cal D}\rho \, e^{-S[\rho]}~~&,&~~
S[\rho] = \int \d x^- \d^2x_\perp \frac{\tr
  \rho(x^-,x_\perp)\, \rho(x^-,x_\perp)} {2\mu^2(x^-)}~.
\label{eq:S_MV}
\eea
Here, $A^+$ is the covariant gauge {\em classical} field (describing the
small-x gluon fields) sourced by the random valence charge density
$\rho$ which one averages over. $\int dx^-\, \mu^2(x^-)$ corresponds
to the average color charge density squared per unit transverse area
and is the only parameter of the model; it is proportional to the
thickness of the nucleus $\sim A^{1/3}$. The expectation value of an
electric Wilson line $V(x_\perp)$, for example, is then computed
as\footnote{$\log\,\langle \tr V\rangle$ is power divergent in the IR
  and so requires a cutoff. We simply write the formal
  Eq.~(\ref{eq:<trV>}) to illustrate the averaging procedure. The
  dipole probe from Eq.~(\ref{eq:<trVV+>}) does not exhibit such a
  power-law divergence in the IR.}
\be \label{eq:<trV>}
\left< \tr V(x_\perp)\right> =
\frac{1}{Z} \int {\cal D}\rho \, e^{-S[\rho]}\, \tr
{\cal P} e^{-ig \int\limits_{-\infty}^\infty dx^- A^+(x^-,x_\perp)}~.
\ee
The forward scattering amplitude ${\cal N}(r)$ of a quark - antiquark
dipole of size $r=|y_\perp-x_\perp|$ is given by
\be \label{eq:<trVV+>}
{\cal N}(r) =
\left< 1 - \frac{1}{N_c} \tr V^\dagger(x_\perp)V(y_\perp)\right> =
\frac{1}{Z} \int {\cal D}\rho \, e^{-S[\rho]}\, \left[1-
  \frac{1}{N_c}\tr
  {\cal P} e^{ig \int\limits_{-\infty}^\infty dx^- A^+(x^-,x_\perp)}
  {\cal P} e^{-ig \int\limits_{-\infty}^\infty dx^- A^+(x^-,y_\perp)}
  \right]~.
\ee
We employ hermitian generators. The size $r$ where ${\cal N}(r)$ grows
to order 1 defines the (inverse) saturation scale $Q_s^{-1}$. In the
MV model one finds that $Q_s^2 \sim C_F g^4 \int dx^- \mu^2(x^-)$. For
transverse momenta $q^2\gg Q_s^2$ the Fourier transform of the forward
scattering amplitude defines the dipole unintegrated gluon
distribution
\be
xG(x,q^2) \simeq g^2 \left< \tr |A^+(q)|^2\right>~.
\ee

Quantum corrections to the MV model modify the statistical weight
$W[\rho] \equiv \exp(-S[\rho])$. Ref.~\cite{Iancu:2002aq} proposed a
Gaussian ``mean-field'' approximation for $W[\rho]$ at small
light-cone momentum fractions (far from the valence sources) which
reproduces the proper gluon distribution (or dipole scattering
amplitude) both at small ($q^2\ll Q_s^2$) as well as at high ($q^2\gg
Q_s^2$) transverse momentum:
\be
W_G[\rho] = e^{-S_G[\rho]}~~,~~
  S_G[\rho] = \int \d^2 x_\perp \d^2 y_\perp \, \frac{\tr
  \rho(x_\perp)\, \rho(y_\perp)} {\mu^2(x_\perp-y_\perp)}~.
\ee
This non-local Gaussian can be rewritten in $q$-space as\footnote{Note
that we define $1/\mu^2(q^2) \equiv \int \d^2 r \, e^{-iqr}/\mu^2(r)$.}
\bea
S_G[\rho] &=& \int \dd q \, \tr \rho(q)\, \rho(-q)
\int \d^2 r \frac{e^{-iqr}}{\mu^2(r)}\notag \\
&\equiv& \int \dd q \frac{\tr \rho(q)\, \rho(-q)}{\mu^2(q^2)}~.
\eea
This action reproduces the correct dipole scattering amplitude and
Weizs\"acker-Williams gluon distribution in the short distance (high
transverse momentum) limit, c.f.\ ref~\cite{Iancu:2002aq}, with
\be \label{eq:mu2_q2}
\mu^2(q^2) \simeq \mu_0^2 \left(\frac{q^2}{Q_s^2}\right)^{1-\gamma}~.
\ee
Here, $\gamma\simeq0.64$ is the BFKL anomalous dimension~\cite{bfkl}
(in the presence of a saturation
boundary~\cite{Mueller:2002zm}). $Q_s^2$ and $\mu_0^2$ are evaluated
at the rapidity of interest (like in the MV model $\mu_0^2$ is again
proportional to the thickness of the nucleus $\sim A^{1/3}$). We will
not spell out this dependence on $Y$ explicitly since our focus here
is not on the growth of $Q_s$ with $Y$ which is well known. For the
present purposes the most important effect of the resummation of
quantum fluctuations is that $\mu^2(q^2)$ increases with transverse
momentum when $q^2>Q_s^2$.

The paper is organized as follows. In Sec.~\ref{Sec:outline} we
present the basic idea for computing an effective potential for a
given observable by introducing a constraint into the functional
integral. In Sec.~\ref{Sec:warmup}, in order to illustrate the
approach with a simple example we compute the effective potential for
the number ${\rm tr}\ \rho^2$ in the MV model on a single site. We
then compute the effective potential for the covariant gauge gluon
distribution function in Sec.~\ref{Sec:effpot}. We proceed to
calculate the fluctuations of the gluon multiplicity and of the
average squared transverse momentum in Sec.~\ref{Sec:glmult}.  In
Sec.~\ref{Sec:MC} we present results of numerical Monte-Carlo
simulations within the MV model and for the solution of the JIMWLK
renormalization group equation. We end with a discussion and
outlook in Sec.~\ref{Sec:conc}.

\section{The basic idea: introducing the constraint effective potential}
\label{Sec:outline}

Expectation values such as those written in
Eqs.~(\ref{eq:<trV>},\ref{eq:<trVV+>})  refer to a statistical
average of an observable $O[\rho]$ over {\em all} configurations
$\rho(x^-,x_\perp)$ from the ensemble $W[\rho]$. On the other hand, we
may be interested in the value of an observable for a specific subset
of configurations such as configurations with a high number of gluons
or with a specific unintegrated gluon distribution. These represent
more ``global'' measures averaging over all fluctuations of
$\rho(x^-,x_\perp)$ which do not affect, say, the unintegrated gluon
distribution. In other words, our goal is to perform the integral over
$\rho$ subject to the contraint that, for example, $O[A^+]=g^2 \tr
|A^+(q)|^2$ is fixed, thereby decomposing the space of all $\rho(q)$,
or $A^+(q)$, into invariant subspaces (w.r.t.\ the given observable).

\begin{figure}[htb]
\begin{center}
\includegraphics[width=0.4\textwidth]{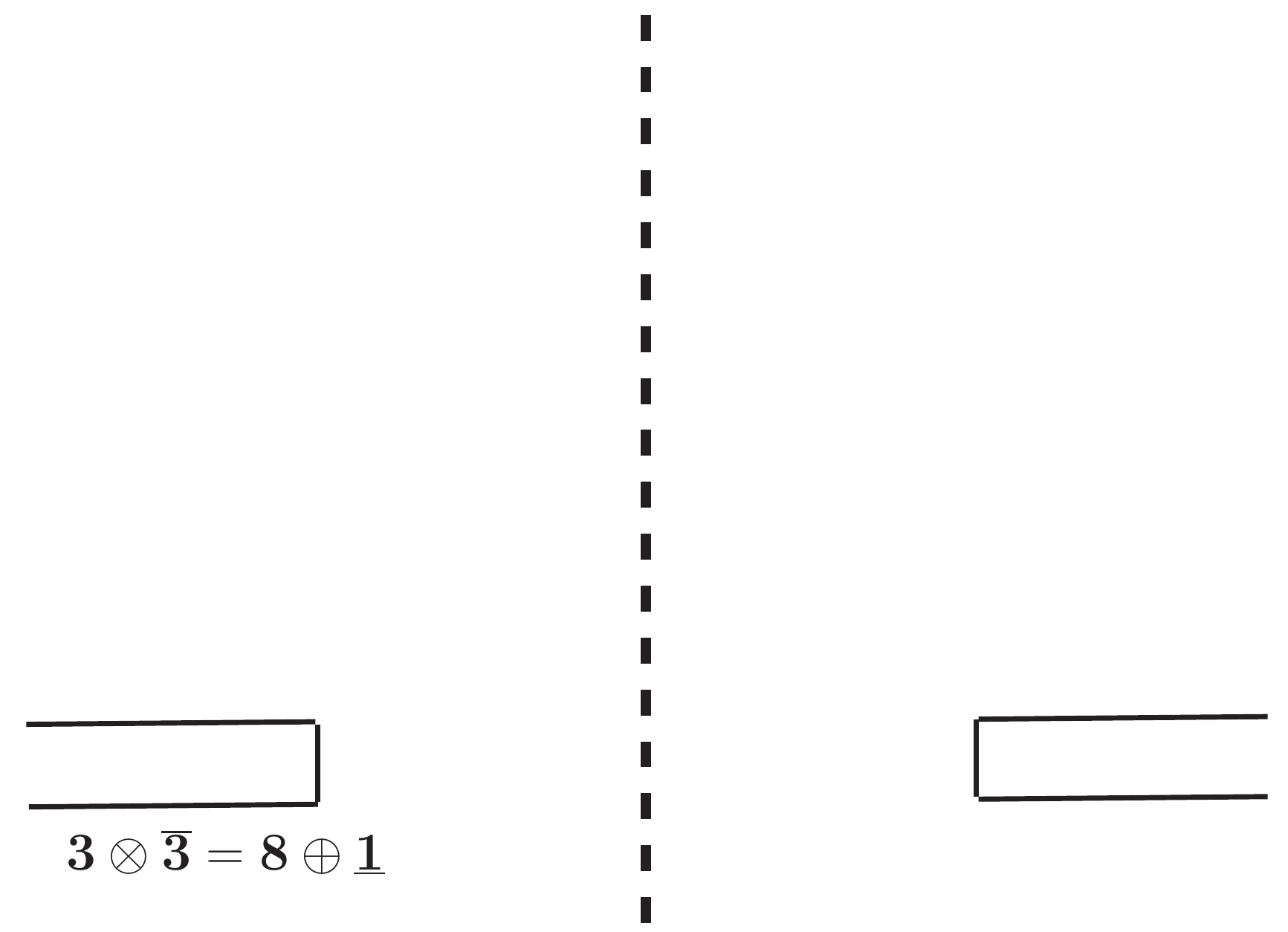}
\hspace*{3cm}
\includegraphics[width=0.4\textwidth]{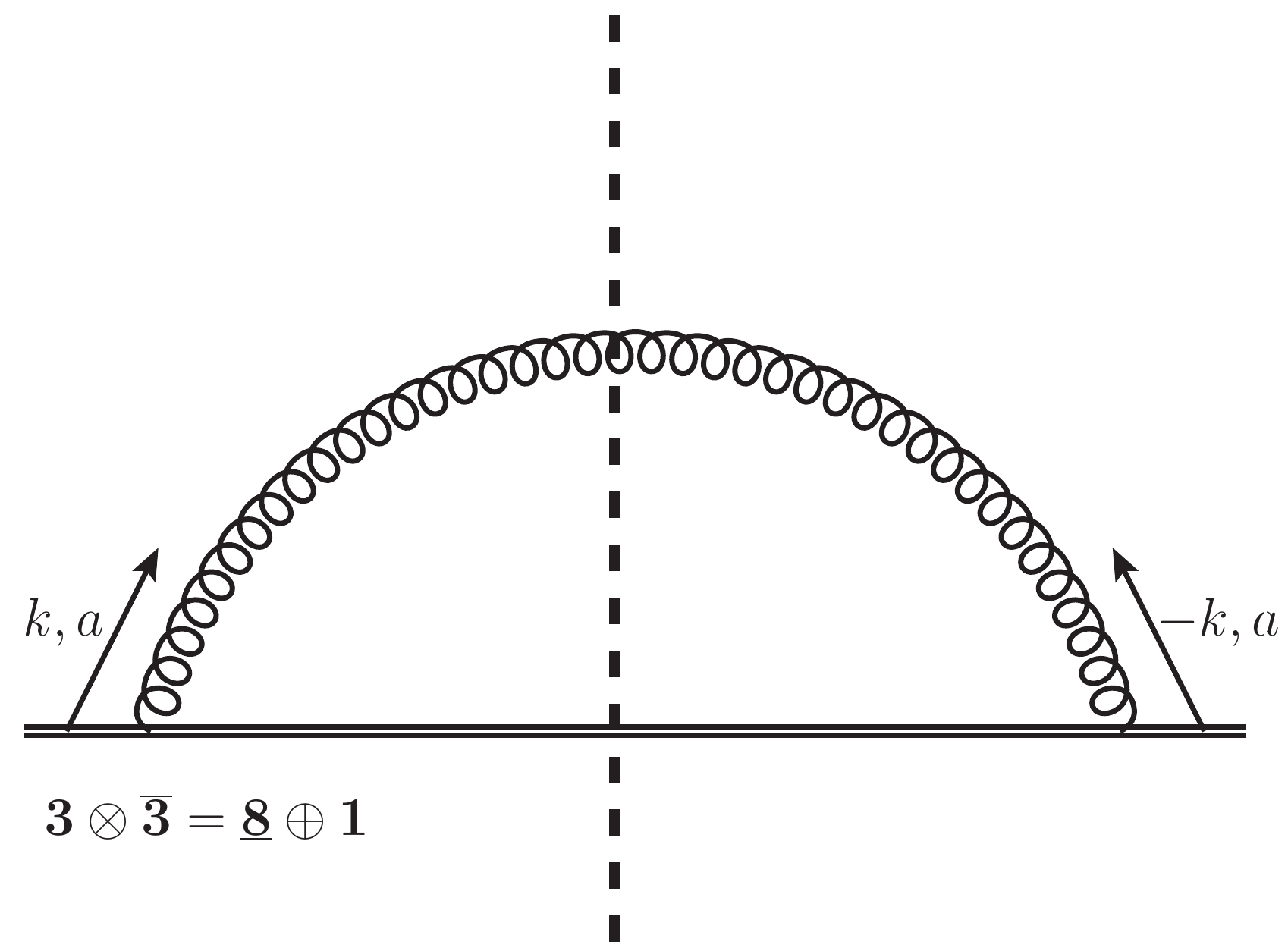}
\end{center}
\caption[a]{Illustration of the fluctuation of the gluon distribution
  $g^2 \tr A^+(k) A^+(-k)$ of a system of a quark and an anti-quark. The
  gluon couples coherently to the charges. Left: if the $q\bar{q}$
  system is in the color singlet state there is no coupling to the
  field and the gluon distribution vanishes. Right: the octet
  representation has a non-vanishing gluon distribution.}
\label{fig:xG_rho_fluc}
\end{figure}
We illustrate the fluctuations of the gluon distribution originating
from the fluctuations of the classical valence color charge density
$\rho$ in Fig.~\ref{fig:xG_rho_fluc}. For simplicity we show a simple
example corresponding to the fluctuations of the color charge
representation of a system composed of a quark and an anti-quark. The
MV model describes the fluctuations of a system of many valence
charges in a high-dimensional representation about the most likely
representation~\cite{JeonVenugopalan,Petreska_rho4}.

The logarithm of the inverse of the partition function obtained after
integrating out the orthogonal fluctuations of $\rho$ then defines an
effective potential\footnote{More generally, this would give the
  effective action for the field $g^2 \tr |A^+(q)|^2$.} for
$O[\rho(q)] = g^2 \tr |A^+(q)|^2$:
\bea
Z &=& \int {\cal D}X(q) \, e^{-V_\mathrm{eff}[X(q)]}~,\label{eq:Z(X)}\\
e^{-V_\mathrm{eff}[X(q)]} &=& \int {\cal D}\rho(q) \, W[\rho(q)] \,
\delta(X(q)-O[\rho(q)])~.
\eea
The stationary point of $V_\mathrm{eff}[X]$ corresponds to the
extremal gluon distribution $X_s(q)$. In the limit of an infinite
number of degrees of freedom, i.e.\ the large-$N_c$ limit in our
case\footnote{To go beyond the large-$N_c$ limit one would have to
  actually compute the integral over $X(q)$ in Eq.~(\ref{eq:Z(X)})
  which can be done by means of a Legendre transformation.}, $X_s(q)$
of course is equal to the expectation value of $\langle g^2 \tr
|A^+(q)|^2\rangle$. Away from the stationary solution, the potential
$V_\mathrm{eff}[X]$ provides insight into the form of fluctuations
about the extremum. Specifically, we shall analyze the correlation
of fluctuations of the number of gluons (above the saturation
scale) and their typical transverse momentum.

Fluctuations of various observables induced by the fluctuations of
$\rho(x_\perp)$ have been analyzed before. For example, the
multiplicity distribution of gluons with transverse momenta above
$Q_s$~\cite{Gelis:2009wh} and the fluctuations of the real and
imaginary parts of spatial Wilson loops~\cite{Lappi:2014opa} in the
central region of a collision of two sheets of color charge have been
analyzed. Angular harmonics of the dipole scattering amplitude ${\cal
  N}(r)$ for random individual configurations $\rho(x_\perp)$ of the
target have been shown in Ref.~\cite{Dumitru:2014vka} (the ensemble
average is, of course, isotropic). The evolution of the imaginary part
of the dipole $S$-matrix, i.e.\ of the odderon $O[A^+]=(-i/2N_c)\tr
[V_x V^\dagger_y - h.c.]$, has been discussed in
Ref.~\cite{Lappi:2016gqe}.  Here, we describe how one can explicitly
integrate out the fluctuations of $\rho$ or $A^+$ under which a given
observable is invariant in order to derive an effective potential for
the observable itself. We apply our method specifically to compute the
effective potential for the covariant gauge gluon distribution $\tr
|A^+(q)|^2$ from which we deduce the correlation of fluctuations of
the multiplicity of hard gluons and of their typical transverse
momentum.
A somewhat similar procedure was previously used to compute the density 
matrix of the soft gluon fields and the associated 
entanglement entropy, see Ref.~\cite{Kovner:2015hga}.

\section{Warm-up: effective potential for $\tr \rho^2$}
\label{Sec:warmup}

We begin with the effective potential for $\tr \rho^2$ on a single
site. The procedure is essentially identical to that used in sec.~III
of Ref.~\cite{Dumitru:2004gd} to compute the effective potential for
Polyakov loops in a single-site matrix model.

The partition function for a single site is
\be \label{eq:Zss}
Z = \int \left(\prod_a \d\rho^a\right) \, e^{-\tr\rho^2 / \mu^2}~.
\ee
$\rho\equiv \rho^at^a$ denotes the random color charge density at the
site and is an element of the algebra of the color group in the
fundamental representation: $\tr t^at^b=\frac{1}{2}\delta^{ab}$.  We
shall keep only contributions to $V_\mathrm{eff}(X)$ of order $N_c^2$
and drop terms of order 1.
 
The goal is to write (\ref{eq:Zss}) in the form
\be
Z = \int \d X e^{-V_\mathrm{eff}(X)}~~~~,~~~~
X \equiv \tr\rho^2 = \frac{1}{2}\rho^a\rho^a ~,
\ee
where $V_\mathrm{eff}(X)$ is the effective potential for $\tr\rho^2$.
In other words, $e^{-V_\mathrm{eff}(X)}$ is the partition sum for
$N_c^2$ scalars $\rho^a$ satisfying the constraint $\rho^a \rho^a =
2X$.

To compute $V_\mathrm{eff}(X)$ we introduce a $\delta$-function
constraint in Eq.~(\ref{eq:Zss}),
\bea \label{eq:Zss_constr}
Z &=& \int \d \lambda \int \left(\prod_a \d\rho^a\right) \, e^{-\tr\rho^2 / \mu^2}
\, \delta\left(\lambda-\tr\rho^2\right) \notag \\
&=& \int \d \lambda \int\frac{\d\omega}{2\pi}
e^{-\frac{\lambda}{\mu^2}-i\omega\lambda}
\underbrace{\int \left(\prod_a \d\rho^a\right) e^{i\omega\;
    \tr\rho^2}}_{\widetilde{Z}(\omega)}~.
\eea
The integral for $\widetilde{Z}(\omega)$ is easily computed in
spherical coordinates,
\bea
\widetilde{Z}(\omega) &=& \int \d X \int \left(\prod_a \d\rho^a\right)
\delta\left(X-\frac{1}{2}\rho^b\rho^b\right)
e^{\frac{1}{2} i\omega \rho^c\rho^c} \notag \\
&\sim& \int \d X \, e^{i\omega X + \frac{1}{2}N_c^2\log X}~.
\eea
In the last step we have dropped an irrelevant $\omega$-independent
normalization factor. This expression for $\widetilde{Z}(\omega)$ then
leads to
\be
Z = \int \d X \, e^{-\frac{X}{\mu^2}+\frac{1}{2}N_c^2\log X}~.
\ee
Hence, the effective potential is given by 
\be
V_\mathrm{eff}(X) = \frac{X}{\mu^2}-\frac{1}{2}N_c^2\log X~.
\ee
The stationary point of this potential is
\be
X_s \equiv \langle\tr\rho^2\rangle = \frac{1}{2} N_c^2 \mu^2~.
\ee
Of course, this result can be obtained directly from the correlator
$\langle \rho^a \rho^b\rangle = \delta^{ab} \mu^2$
which follows from the action in Eq.~(\ref{eq:Zss}).

\section{Effective potential for the gluon distribution}
\label{Sec:effpot}

In this section we compute the effective potential for the gluon
distribution $\tr |A^+(q)|^2$ obtained from the field in covariant
gauge. We will also comment briefly on the potential for the gluon
distribution obtained from the light-cone gauge field $\tr |A^i(q)|^2$.

It is convenient to work directly in momentum space. The
partition function of the Gaussian model is taken as
\be \label{eq:Z_rho_q}
Z = \int \left(\prod_q \prod_a \d\rho^a_q\right) e^{-S[\rho]}~~,~~
S[\rho] = \int \dd q \frac{\tr|\rho_q|^2}{\mu^2(q)}~.
\ee
The constraint $(\rho_q)^* = \rho_{-q}$ is implicit. In general, $\mu^2(q)$
may depend on the transverse momentum which would correspond
to a non-local Gaussian action in coordinate space; this dependence is
also left implicit from now on since it is not essential for the
following steps.

Our goal is to obtain an expression of the form
\be
Z = \int {\cal D} X(q) \, e^{-V_\mathrm{eff}[X(q)]}
\ee
for
\be \label{eq:X_intd2b_A+A+}
X(q) \equiv g^2 \tr |A^+(q)|^2 =
\int \d^2b\, \d^2r \, e^{-iq\cdot r} \, g^2 \tr A^+(x_\perp) A^+(y_\perp)~,
\ee
where $b=(x_\perp+y_\perp)/2$ and $r=y_\perp-x_\perp$. Since
$A^+(q)=(g/q^2)\, \rho(q)$ we can write the partition sum as
\bea
Z &=& \prod_q \int \d\lambda_q\frac{\d\omega_q}{2\pi} \left(\prod_a
\d\rho^a_q\right) e^{-i\omega_q\lambda_q +i\omega_q\frac{g^4}{q^4}\tr
  |\rho_q|^2}
e^{-\dd q \frac{q^4}{g^4} \frac{\lambda_q}{\mu^2}} \notag \\
&=& \left[
  \prod_q \int \d\lambda_q\frac{\d\omega_q}{2\pi}
  e^{-i\omega_q\lambda_q} e^{-\dd q \frac{q^4}{g^4} \frac{\lambda_q}{\mu^2}}
  \right]
\underbrace{\prod_q \int \left(\prod_a\d\rho^a_q\right)
  e^{i\omega_q\frac{g^4}{q^4}\tr|\rho_q|^2}}_{\widetilde{Z}[\omega_q]}~. \label{eq:Z_omega_q}
\eea
The first line of the equation above is obtained from the original
partition sum~(\ref{eq:Z_rho_q}) by inserting a $\delta$-functional
\be
1 = \int \prod_q \d\lambda_q \, \delta\left(\lambda_q-\frac{g^4}{q^4}\tr
|\rho_q|^2 \right)
\ee
which fixes $\lambda(q)=(g^4/q^4)\tr |\rho(q)|^2$.
To compute $\widetilde{Z}(\omega_q)$ we again introduce the constraint
field $g^2\tr|A^+(q)|^2=X(q)$,
\bea
\widetilde{Z}[\omega_q] &=& \int \prod_q \d X_q
\left(\prod_a \d\rho^a_q\right)
\delta\left(X_q - \frac{g^4}{q^4}\tr|\rho_q|^2 \right)
e^{i\omega_q \frac{g^4}{q^4} \tr|\rho_q|^2} \notag \\
&\sim& \prod_q \int \d X_q \,
X_q^{\frac{N_c^2}{2}} e^{i\omega_q X_q}~.
\eea
Inserting this into Eq.~(\ref{eq:Z_omega_q}) we obtain
\bea
Z &=& \prod_q \int \d X_q \, e^{-\dd q \frac{q^4}{g^4}
  \frac{X_q}{\mu^2} +\frac{1}{2}N_c^2\log X_q}\notag \\
&=& \int {\cal D}X(q) \, e^{-\int \dd q \left[\frac{q^4}{g^4\mu^2}X(q) -
  \frac{1}{2}A_\perp N_c^2\log X(q)\right]}~.
\eea
In the last step we have taken the continuum limit, $A_\perp$ is the
transverse area covered by the integration over the impact parameter
$b$ in Eq.~(\ref{eq:X_intd2b_A+A+}). The effective
potential for the function $X(q)$ is therefore
\be \label{eq:Veff_A+A+}
V_\mathrm{eff}[X(q)] = \int \dd q \left[\frac{q^4}{g^4\mu^2}X(q) -
  \frac{1}{2}A_\perp N_c^2\log X(q)\right]~,
\ee
and its stationary point corresponds to the average gluon distribution
(at order $\sim N_c^2$)
\be \label{eq:Xs_A+A+}
\frac{\delta}{\delta X(k)} V_\mathrm{eff}[X(q)] = 0~ \rightarrow~
X_s(k) \equiv \langle g^2 \tr |A^+(k)|^2\rangle
= \frac{1}{2} N_c^2 A_\perp \frac{g^4\mu^2}{k^4}~.
\ee
The contribution to $V_\mathrm{eff}[X(q)]$ at zeroth order in $N_c$
can be restored by comparing to $X_s(k)$ obtained directly from the
Gaussian two-point function:
\be \label{eq:rho_rho_Xs}
\left< \rho^a(k)\, \rho^b(q)\right> = \delta^{ab} \, (2\pi)^2 \delta(k+q)
\, \mu^2 ~~\rightarrow ~~
X_s(k) = \frac{1}{2} \left(N_c^2-1\right) A_\perp \frac{g^4\mu^2}{k^4}~.
\ee
Hence, to restore the ${\cal O}(1)$ contribution to
$V_\mathrm{eff}[X(q)]$ Eq.~(\ref{eq:Veff_A+A+}) should be modified to
\be \label{eq:Veff_A+A+_allNc}
V_\mathrm{eff}[X(q)] = \int \dd q \left[\frac{q^4}{g^4\mu^2}X(q) -
  \frac{1}{2}A_\perp \left(N_c^2-1\right)\log X(q)\right]~.
\ee
We shall mostly focus on Eq.~(\ref{eq:Veff_A+A+}) in what follows but
refer to~(\ref{eq:Veff_A+A+_allNc}) when ${\cal O}(N_c^0)$ accuracy of
the gluon distribution $X(q)$ would be needed.

The two-point correlator $\langle \rho^a(q)\rho^b(k)\rangle$ of the
color charge, averaged over {\em all} of its fluctuations, follows
from the action $S[\rho]$ and is written in Eq.~(\ref{eq:rho_rho_Xs}).
In order to explicitly split off the fluctuations of $\rho(q)$ at
fixed $X(q)$ we can write this in the form
\be
\langle \rho^a(q)\rho^b(k)\rangle = 2\frac{q^4}{g^4}
\frac{1}{N_c^2A_\perp} \delta^{ab}\, (2\pi)^2\,
\delta(k+q) \int {\cal D}X(l) \, e^{-V_\mathrm{eff}[X]}\, X(q)~.
\ee
Replacing the integration over $X(q)$ by the extremal solution
$X_s(q)$ reproduces the correlator from Eq.~(\ref{eq:rho_rho_Xs}). This last
expression should be useful for future applications where one may want
to explicitly isolate the fluctuations of the gluon distribution $X(q)$
from more complicated expressions involving two-point functions of $\rho(q)$.

We briefly pause our derivation at this point to comment on the
potential describing fluctuations of the Weizs\"acker-Williams gluon
distribution defined via the light-cone gauge field $A^i(q)$. Because
of the non-linear dependence of $A^i$ on $\rho$ we are unable to
compute the effective potential analytically except in the weak field
regime where $A^i(q) = ig(q^i/q^2)\rho(q)$. Hence, in this regime both
the diagonal as well as the off-diagonal components of the WW gluon
distribution, $\delta^{ij} \tr A^i(q) A^j(-q)$ and
$(2q^iq^j/q^2-\delta^{ij})\tr A^i(q) A^j(-q)$, respectively, are equal
to $q^2 \tr A^+(q)A^+(-q)$. The effective potential for these
distributions is therefore again given by Eq.~(\ref{eq:Veff_A+A+})
with the replacement $q^4\to q^2$ in the first term of the integrand.

As a second aside we briefly illustrate the modifications due to adding a
quartic color charge operator to the quadratic action. We choose a
particularly simple form in order to be able to compute the effective
potential exactly without having to resort to a perturbative expansion:
\be
S_4 = \frac{1}{\beta} \int \d^2x \d^2y\, 
\rho^a(x)\rho^a(x)\rho^b(y)\rho^b(y)
= \frac{1}{\beta} \int\dd{q_1}\dd{q_2}\, \rho^a(q_1)\rho^a(-q_1)
\rho^b(q_2)\rho^b(-q_2)~.
\ee
This replaces eq.~(\ref{eq:Z_omega_q}) by
\bea
Z &=& \left[
  \prod_q \int \d\lambda_q\frac{\d\omega_q}{2\pi}
  e^{-i\omega_q\lambda_q}   \right]
e^{-\dd q \sum_q \frac{q^4}{g^4}
    \frac{\lambda_q}{\mu^2} - \dd q\dd q \frac{1}{\beta\, g^4}
    \sum_{q_1,q_2} q_1^4 q_2^4 \lambda(q_1) \lambda(q_2)}
{\widetilde{Z}[\omega_q]} \\
&=&
\int {\cal D}X(q) \, e^{-\int \dd q \left[\frac{q^4}{g^4\mu^2}X(q) 
    -  \frac{1}{2}A_\perp N_c^2\log X(q)\right]
 - \frac{1}{\beta\, g^4}\left(\int \dd q q^4X(q)\right)^2
}
~.\label{eq:Z4_omega_q}
\eea
Hence, in this case
\be \label{eq:Veff4_A+A+}
V_\mathrm{eff}[X(q)] = \int \dd q \left[\frac{q^4}{g^4\mu^2}X(q) -
  \frac{1}{2}A_\perp N_c^2\log X(q)\right]
+ \frac{1}{\beta\, g^4}\left(\int \dd q q^4X(q)\right)^2 ~.
\ee
In the MV model $\mu^2\sim g^2 A^{1/3}$, where $A^{1/3}$ denotes the
thickness of the nucleus, while the coupling $\beta$ for the quartic
color charge density operator involves two additional powers of $g
A^{1/3}\gg1$~\cite{Petreska_rho4}.  Such a quartic in $\rho$ operator
therefore represents a higher order correction in the high gluon
density power counting scheme where $g^4 A^{1/3}={\cal O}(1)$,
c.f.\ next subsection. Moreover, in fig.~\ref{fig:Veff} below we shall
show that the exact numerical solution of the LO small-$x$ evolution
equation agrees rather well with the effective potential for the gluon
distribution derived from a quadratic action. We will therefore
neglect $S_4$ in what follows.

We now return to our discussion of the fluctuations of $X(q)=g^2 \tr
|A^+(q)|^2$ in the model with a quadratic action and write
\be
X(q) = X_s(q) + \delta X(q)
\ee
and expand $V_\mathrm{eff}[X(q)]$ to quadratic order in $\delta
X(q)$. This ``one loop'' approximation leads to
\bea
\Delta V_\mathrm{eff}[\delta X(q)] &\equiv& V_\mathrm{eff}[X(q)]
- V_\mathrm{eff}[X_s(q)] \notag  \\
&\simeq& \frac{1}{2} \int \dd l \dd k \, \delta X(l)
\left\{
\frac{\delta}{\delta X(l)}\frac{\delta}{\delta X(k)}
V_\mathrm{eff}[\delta X(q)]\right\} \delta X(k) \notag \\
&=& \frac{1}{2} \int \dd q \frac{\delta X(q)^2}{X_s(q)^2}
\frac{1}{2} N_c^2 A_\perp \\
\rightarrow \int {\cal D} \, \delta X(q) ~e^{-V_\mathrm{eff}[\delta X(q)]}
&=& e^{-\frac{1}{2}\tr\log \left(\frac{1}{2} 
  \frac{N_c^2 A_\perp}{X_s(q)^2}\right)}~.
\eea

However, it is clear from the form of $V_\mathrm{eff}[X(q)]$ that the
quadratic approximation can not describe fluctuations far from
the extremal solution $X_s(q)$. We therefore follow a different
route. We introduce the fluctuation field $\eta(q)$ through
\be \label{eq:X_Xs_eta}
X(q) = X_s(q)\, \eta(q)~,
\ee
with $X_s(q)$ as written in Eq.~(\ref{eq:Xs_A+A+}). A fluctuation from
the extremal ``path'' $X_s(q)$ has action
\bea
\Delta V_\mathrm{eff}[\eta(q)] &\equiv& V_\mathrm{eff}[\eta(q)]
- V_\mathrm{eff}[\eta(q)=1]\notag  \\
&=&
\frac{1}{2} N_c^2 A_\perp \int \dd q \left[ \eta(q) -1 -
  \log\eta(q)\right]~.
\label{eq:Veff_eta}
\eea
This is a Liouville action (without kinetic term and with negative
Ricci scalar) for the field $\phi(q)=\log\eta(q)$ in two dimensional
$q$-space\footnote{In Ref.~\cite{Iancu:2007st} Iancu and McLerran
  proposed a Liouville action to describe the fluctuations of $Q_s$ in
  the transverse impact parameter plane ($x$-space) due to stochastic
  high-energy evolution; this is unrelated to our discussion of
  fluctuations in the ensemble of gluon distributions $X(q)$ which
  occur even at fixed $Q_s$, as considered here, and exist even in the
  absence of QCD evolution (MV model).}. Indeed, the canonical
dimension of the fluctuation field $\eta(q)$ as introduced in
Eq.~(\ref{eq:X_Xs_eta}) is zero. This will become important below to
understand the spectrum of fluctuations from small-$x$ evolution.

In the following section we use expression~(\ref{eq:Veff_eta}) to
analyze the correlation of gluon number and transverse momentum
fluctuations.

\subsection{Parametric dependence on the number of colors and on the
  thickness of the target} \label{sec:parametric_Nc_A13}

In this subsection we discuss the parametric dependence of the
fluctuations on $N_c$ and on the thickness of the target nucleus which
is proportional to the third root of its atomic number, $A^{1/3}$.  In
particular, we outline that the fluctuations of the gluon distribution
considered here are of the same order in $A^{1/3}$ as the ``extremal''
(or average) gluon distribution $X_s(q)$, and of the same or lower
order in $N_c$. As explained by Kovchegov~\cite{Kovchegov:1999ua},
quantum evolution at leading order applies when $\alpha_s\ll1$ with
$\alpha_s^2 A^{1/3}\sim1$. The latter condition implies that
contributions which do not exhibit longitudinal coherence, i.e.\ those
which are not proportional to the thickness of the nucleus, in this
power counting scheme formally correspond to higher order corrections.

Recall from the previous section that the average gluon distribution
$X_s(q) \sim N_c^2 g^4\mu^2 \sim N_c^2 \alpha_s^2 A^{1/3}$. The {\em
  action} (\ref{eq:Veff_A+A+}) evaluated at $X_s(q)$ is
$V_\mathrm{eff}[X_s(q)] \sim N_c^2$ (times a numerical factor equal to
zero in dimensional regularization in $D=2-\epsilon$ dimensions). This
corresponds to the action of classical gluon fields times a factor of
$g^2$ from the coupling to the sources (see
Fig.~\ref{fig:xG_rho_fluc}).

In order to be able to evolve initial fluctuations to small $x$ using
leading order evolution these fluctuations $\delta X(q)\equiv X_s(q)\,
\eta(q)$ must also be of order $A^{1/3}$. This is satisfied since the
effective action~(\ref{eq:Veff_eta}) for the fluctuation field
$\eta(q)$ does not involve the thickness $\sim A^{1/3}$
explicitly. Indeed, the MV model~\cite{MV} outlined in the
Introduction describes precisely these longitudinally coherent valence
color charge fluctuations. In other words, fluctuations $\delta
X(q)\equiv X_s(q)\,\eta(q)$ corresponding to a penalty action $\Delta
V_\mathrm{eff}[\eta(q)]$ which is {\em independent} of $A^{1/3}$ are
of the same order in $A^{1/3}$ as the average gluon distribution
$X_s(q)$ and can be evolved to small $x$. However, one can not study
fluctuations with a suppression probability $p$ such that
$\log p^{-1}=V_\mathrm{eff}[\eta(q)]\sim (A^{1/3})^{-1}$ since that
would correspond to $\eta(q) = {\cal O}((A^{1/3})^{-1})$ and $\delta
X(q) = {\cal O}((A^{1/3})^0)$. Such fluctuations are of higher order
in the coupling~\cite{Kovchegov:1999ua}.

Power counting in $N_c$ proceeds along similar lines. $\Delta
V_\mathrm{eff}[\eta(q)]$ is explicitly proportional to $N_c^2$, so
$\eta(q) = {\cal O}(N_c^0)$ corresponds to fluctuations $\delta X(q)$
at the same order in $N_c$ as the average gluon distribution $X_s(q)$.
These can be selected by an external ``trigger'' probability $p$ such
that $\log p^{-1}=V_\mathrm{eff}[\eta(q)]\sim N_c^2$. However, it is
allowed to select less suppressed fluctuations corresponding to $\log
p^{-1} \sim N_c^0$ provided such terms in the effective action are
accounted for, c.f.\ Eq.~(\ref{eq:Veff_A+A+_allNc}).

\section{Gluon multiplicity and transverse momentum fluctuations}
\label{Sec:glmult}

In this section we analyze fluctuations of the semi-hard gluons above
the saturation momentum $Q_s$ and up to a maximum momentum scale
$Q_\mathrm{max} \gg Q_s$. The number of such gluons for a given
$X(q)=g^2\tr |A^+(q)|^2$ is given by
\be
N_g[X(q)] = \int \dd q q^2 X(q) = \int \dd q q^2 X_s(q)\, \eta(q)~.
\ee
The integral extends from $q^2=Q_s^2$ up to $q^2=Q_\mathrm{max}^2$. As
already mentioned above in the linear regime $q^2\, g^2 \tr
|A^+(q)|^2$ approaches the Weizs\"acker-Williams gluon distribution
$g^2 \tr |A^i(q)|^2$, so $N_g$ counts the number of
Weizs\"acker-Williams gluons from $Q_s^2$ to $Q_\mathrm{max}^2$. We
focus first on the MV model with $\mu^2$=const; analogous results for
a $q$-dependent $\mu^2(q)$ shall be summarized at the end of this
section.

The number of additional gluons due to the fluctuation about the
extremal gluon distribution is given by
\be \label{eq:DNg}
\Delta N_g[\eta(q)] = \int \dd q q^2 X_s(q)\, \left[\eta(q)-1\right]~.
\ee
This quantity does not depend on the UV cutoff $Q_\mathrm{max}^2$
because the fluctuation has finite support in order to have a finite
action.

The average (squared) transverse momentum of gluons between
$q^2=Q_s^2$ and $q^2=Q_\mathrm{max}^2$ can be defined through (see
analogous discussion in Ref.~\cite{Baier:1996sk})
\be
\overline{q^2}[X(q)] = \frac{\int \dd q q^2 X(q)}{\int \dd q X(q)} =
\frac{N_g[X(q)]}{\int \dd q X(q)}~.
\ee
Here $\overline{q^2}$ refers to an average over the
transverse momentum distribution for a given gluon distribution $X(q)$
but {\em not} to an average over all configurations of $A^+$. Once
again we subtract the value at the saddle point,
\be
\Delta \overline{q^2}[\eta(q)] = 
\frac{\Delta N_g[\eta(q)]}{\int \dd q X_s(q)\eta(q)}~.
\ee
We now proceed to discuss the effect of fluctuations,
$\eta(q)\neq1$. Our strategy is to introduce a trial function for
$\eta(q)$ for which we then evaluate $N_g$, $\overline{q^2}$, and the
``penalty action'' $\Delta S$ via Eq.~(\ref{eq:Veff_eta}).  Consider
the {\em ansatz}
\be \label{eq:eta(q)}
\eta(q) = 1 + \eta_0 \left(\frac{g^4\mu^2}{q^2}\right)^a\,
\Theta\left(q^2-\Lambda^2\right)
\Theta\left(Q^2-q^2\right)
~.
\ee
Thus, the fluctuation has support on the interval $\Lambda^2<q^2<Q^2$
within the window $Q_s^2<q^2<Q_\mathrm{max}^2$, i.e.\ $\Lambda^2\ge
Q_s^2$, $Q^2\le Q_\mathrm{max}^2$ with $\Lambda^2 \ll Q^2$. Also, by
dimensional analysis the multiplicative fluctuation can depend only on
$q^2/\mu^2$ since $\mu^2$ is the only dimensionful scale in the MV
action~(\ref{eq:S_MV}). We recall from our discussion in
sec.~\ref{sec:parametric_Nc_A13} that, parametrically,
\be
\eta_0 \sim \frac{1}{N_c^2 \, (g^4\mu^2)^{a}} \, \Delta S ~,
\ee
so that for $\Delta S\sim N_c^2$, $\delta X(q) = X_s(q)\,\eta(q)$ is
of the same order in $N_c$ and $A^{1/3}$ as the average gluon
distribution $X_s(q)$.

For a fluctuation of the form~(\ref{eq:eta(q)}) the excess gluon
multiplicity is given by
\be \label{eq:MV_dNg}
\Delta N_g \simeq \frac{1}{8\pi} N_c^2 A_\perp g^4\mu^2
\eta_0\times
\begin{cases}
  \frac{1}{a}\left(\frac{g^4\mu^2}{\Lambda^2}\right)^{a} & (a>0)~, \\
  \log\frac{Q^2}{\Lambda^2} & (a=0)~,\\
 \frac{1}{|a|}\left(\frac{Q^2}{g^4\mu^2}\right)^{|a|} & (a<0)~.
\end{cases}
\ee
The excess (squared) transverse momentum of gluons with transverse
momentum above the saturation scale is given by
\be \label{eq:qbar2}
\Delta \overline{q^2}\simeq
Q_s^2\,\eta_0\times
\begin{cases}
  \frac{1}{a}\left(\frac{g^4\mu^2}{\Lambda^2}\right)^{a} & (a>0)~, \\
  \log\frac{Q^2}{\Lambda^2} & (a=0)~,\\
 \frac{1}{|a|}\left(\frac{Q^2}{g^4\mu^2}\right)^{|a|} & (a<0)~,
\end{cases}
\ee
or
\be \label{eq:Ng_q2}
\Delta N_g \simeq N_c A_\perp \, \Delta \overline{q^2} ~.
\ee
We have simplified the expression by linearizing in the fluctuation
amplitude.  The factor of $N_c$ in this equation arises due to the
fact that we only integrate over gluons with $q^2>Q_s^2$ with $Q_s^2
\sim N_c \, g^4\mu^2$. According to Eq.~(\ref{eq:Ng_q2}) the average
squared transverse momentum due to the fluctuation is proportional to
the excess number of gluons it contains. In the next section we shall
confirm such a tight nearly linear correlation of $\Delta N_g$ and
$\Delta \overline{q^2}$ via Monte-Carlo simulations.

Finally, the penalty action for such a fluctuation $\eta(q)$ is
\be \label{eq:Delta_S_eta}
\Delta S[\eta(q)] \simeq \frac{1}{8\pi} N_c^2 A_\perp \, g^4\mu^2\, \eta_0
\times
\begin{cases}
  \frac{1}{1-a} \left(\frac{Q^2}{g^4\mu^2}\right)^{1-a} & (a<1)~, \\
  \log\frac{Q^2}{\Lambda^2} & (a=1)~,\\
 \frac{1}{a-1}\left(\frac{g^4\mu^2}{\Lambda^2}\right)^{a-1} & (a>1)~.
\end{cases}
\ee
The goal now is to pay as low a price $\Delta S[\eta(q)]$ as possible
while maximizing $\Delta N_g$ and $\Delta \overline{q^2}$.
Fluctuations with $a<0$, corresponding to increasing $\eta(q)$, come
with a large penalty $\Delta S$. In fact, even a flat $\eta(q)$ with
$a\to0$ corresponds to $\Delta S\sim Q^2$ while, at the same time,
$\Delta N_g$ and $\Delta \overline{q^2}$ increase only logarithmically
with $Q^2$. Similarly, fluctuations with $a>1$, which drop off very
rapidly with $q^2$, give small $\Delta S$, but also a small
multiplicity excess $\Delta N_g$.  Therefore, we expect that in the MV
model the dominant ``high multiplicity'' fluctuations would have a
high-$q$ tail corresponding to $1>a>0$.

We now turn to a $q$-dependent $\mu^2(q)$ as written in
Eq.~(\ref{eq:mu2_q2}). This corresponds to the non-local Gaussian
approximation to the JIMWLK action at small $x$ proposed in
Ref.~\cite{Iancu:2002aq} which accounts for the small-$x$ anomalous
dimension.  Here, the gluon excess above $Q_s^2$ is
\be \label{eq:IIM_dNg}
\Delta N_g[\eta(q)] \simeq \frac{1}{8\pi} N_c^2 A_\perp g^4\mu_0^2
    \frac{\eta_0}{1-\gamma-a}
    \left(\frac{Q^2}{Q_s^2}\right)^{1-\gamma}
    \left(\frac{g^4\mu_0^2}{Q^2}\right)^a
    ~,~~~~~~~(1-\gamma >a)
\ee
while the additional transverse momentum contributed by the
fluctuation is
\be
\Delta \overline{q^2}[\eta(q)] \simeq
     Q_s^2\frac{\eta_0}{1-\gamma-a}
     \left(\frac{Q^2}{Q_s^2}\right)^{1-\gamma}
     \left(\frac{g^4\mu_0^2}{Q^2}\right)^a
~~~~~~~~~      (-\gamma <a<1-\gamma)~.
\ee
Once again we have linearized this expression in $\eta_0$.
In this approximation the proportionality~(\ref{eq:Ng_q2}) of $\Delta N_g$ and
$\Delta \overline{q^2}$ still holds.

The ``penalty'' action for a fluctuation $\eta(q)\ne1$ is again given
by Eq.~(\ref{eq:Delta_S_eta}) with $\mu^2\to\mu_0^2$. Contrary to the
MV model, near scale invariant fluctuations with $a\approx0$ may now
be significant. While they do come with a ``penalty'' proportional to
$Q^2$ ($\Delta S \sim \eta_0 N_c^2 A_\perp Q^2$) they also increase
substantially the gluon number $\Delta N_g$ and the transverse
momentum $\Delta \overline{q^2}$ by a power rather than a logarithm of
$Q^2$.

\section{Monte-Carlo simulations}
\label{Sec:MC}

In this section we show results of numerical Monte-Carlo
simulations. The technical aspects of these Monte-Carlo simulations
are standard by now, our specific implementation has been discussed in
some detail in Ref.~\cite{Dumitru:2014vka}. We generate random
color charge configurations according to the MV model action; for each
configuration we have computed the number of gluons $N_g$ as well as
their average (squared) transverse momentum $\overline{q^2}$ as
described in the previous section. These quantities have been
integrated up to the lattice cutoff at about $Q\sim 85Q_s$ (for $Y=0$)
resp.\ $Q\sim 35Q_s$ (for $\alpha_s Y=1$). We should stress that these
initial configurations have been generated with a uniform $\mu^2$
across the transverse impact parameter plane. Hence, there are no
``voids'' in the target nor is there a boundary to vacuum.

We have also solved the leading order B-JIMWLK renormalization
group equation~\cite{balitsky,jimwlk} at fixed coupling to a rapidity
$Y=1/\alpha_s$ and performed a similar analysis on those
configurations. JIMWLK evolves each Wilson line $V(x_\perp)$ from
rapidity 0 to $Y$ where we take
\be \label{eq:X_q}
X(q) = \int \d^2b \int \d^2r \, e^{-i qr} \, \tr
\left[V_Y\left(b-\frac{r}{2}\right) 
V_Y^\dagger\left(b+\frac{r}{2}\right) -1\right]~.
\ee
The saturation scale $Q_s(Y)$ is determined implicitly from the dipole
forward scattering amplitude introduced in Eq.~(\ref{eq:<trVV+>})
above: ${\cal N}_Y(r=\sqrt{2}/Q_s) = 1-1/\sqrt{e}$. Note that ${\cal
  N}_Y(r)$ is averaged over all configurations.

The gluon distribution at a fixed impact parameter $b$ is given by the
Wigner distribution
\be
X_\mathrm{W}(q,b) \equiv \frac{\d X(q)}{\d^2 b} =
\int \d^2r \, e^{-i qr} \, \tr
\left[V_Y\left(b-\frac{r}{2}\right) 
V_Y^\dagger\left(b+\frac{r}{2}\right) -1\right]~.
\ee
$X_\mathrm{W}(q,b)$ is neither real (for $N_c\ge3$ colors) nor
positive definite since gluons can not be localized
both in transverse momentum and impact parameter space. The Wigner
distribution has to be averaged over transverse area patches of linear
dimension $\gsim 1/Q_s$ to be interpreted as the distribution of
gluons with transverse momenta $q\ge Q_s$.

Computationally instead it is much more efficient to analyze
\be \label{eq:X_R_q_b}
X_R(q,b) = \int \d^2x \int \d^2y \, e^{-i q(y-x)} \, 
e^{-\frac{(b-x)^2}{2R^2}}
e^{-\frac{(b-y)^2}{2R^2}}
\tr \left[V_Y(x) V_Y^\dagger(y) -1\right]~,
\ee
which is similar to the smeared Wigner distribution of hard gluons.
It corresponds to the gluon distribution at impact parameter $b$
averaged over distance scales of order $R$. The limit $R\to\infty$
takes $X_R(q,b)$ back to $X(q)$. The numerical results
presented below were obtained using $R=2/Q_s(Y)$.

The B-JIMWLK equations describe fluctuations only up to scales where
the dipole scattering amplitude drops to ${\cal O}(\alpha_s^2)$, see
the review~\cite{EvolFlucs} and references therein. At such scales the
fact that the number of gluons in the hadron is discrete leads to
large fluctuations in the evolution
speed~\cite{Hatta:2006hs}. However, the running of the coupling in QCD
delays the effects of these fluctuations (related to the discrete
number of gluons) to very high
rapidities~\cite{Dumitru:2007ew}. Hence, for rapidities and transverse
momenta of practical interest the JIMWLK equations may be a useful
approximation, at least for those regions in impact parameter space
where the gluon density is not too low.

We now present the results obtained from the Monte-Carlo
simulation.
\begin{figure}[htb]
\begin{center}
\includegraphics[width=0.49\textwidth]{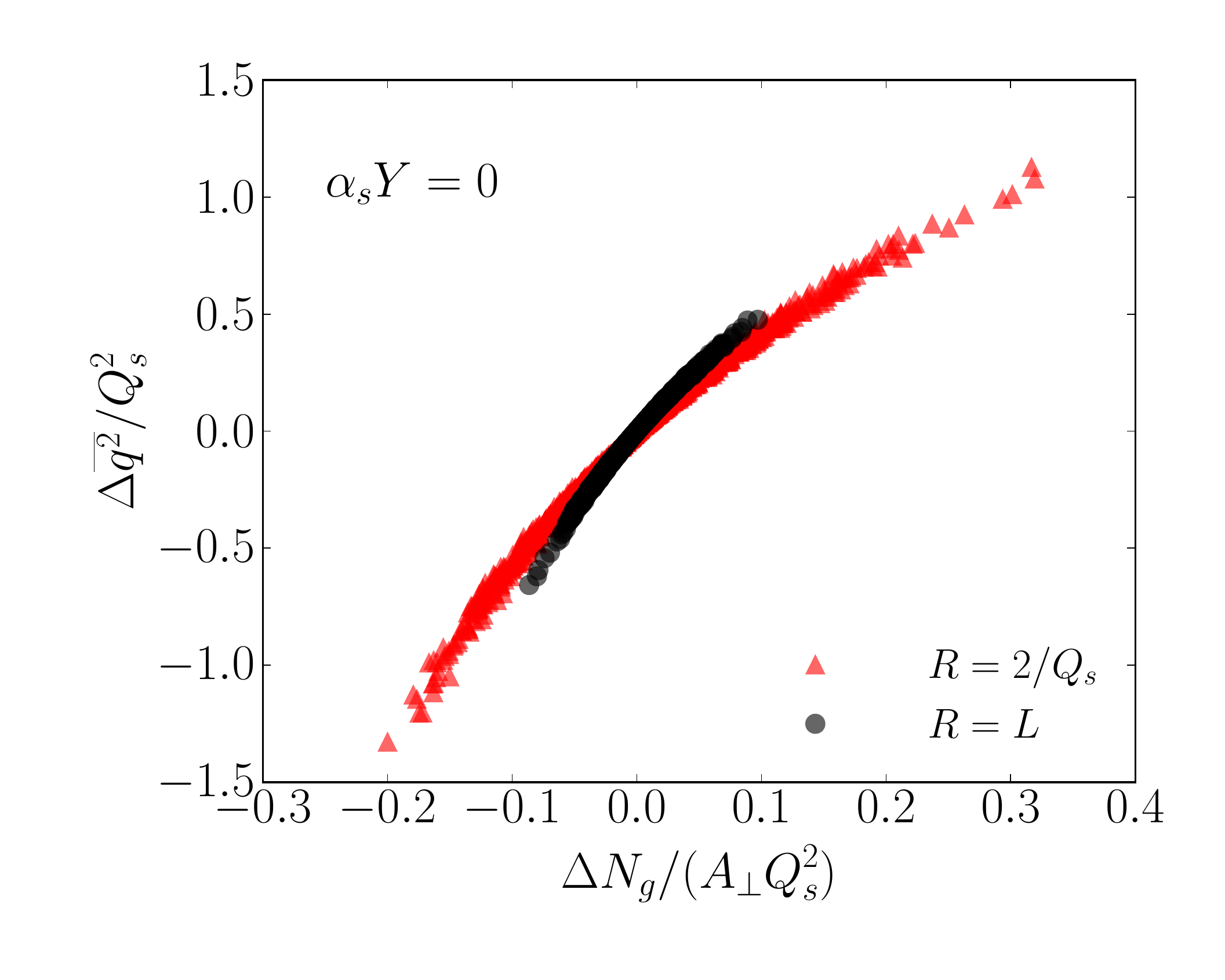}
\includegraphics[width=0.49\textwidth]{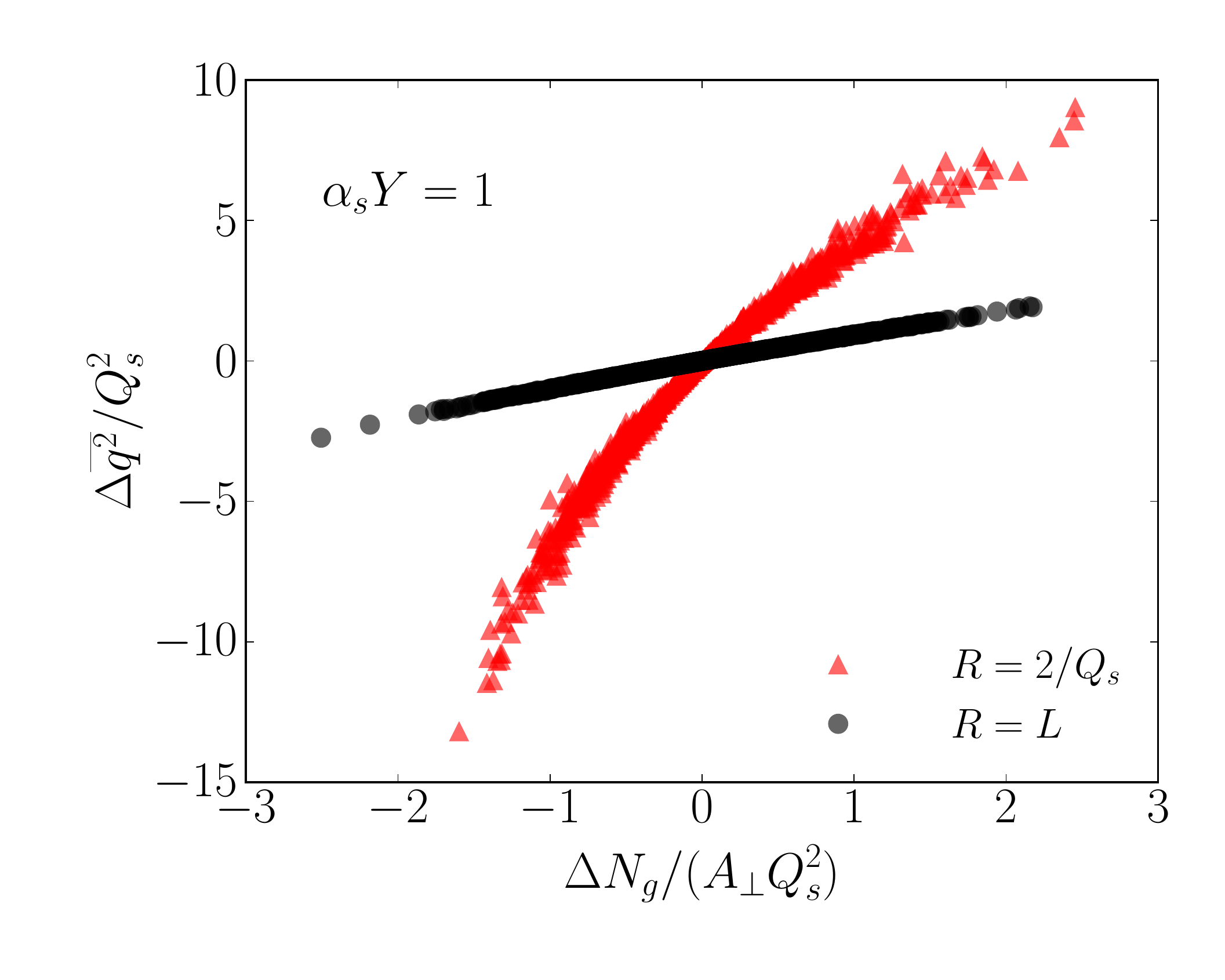}
\end{center}
\vspace*{-1cm}
\caption[a]{Fluctuations of the gluon density and average squared
  transverse momentum in a random Monte-Carlo sample of 150
  configurations of the small-$x$ fields. The gluon distribution has
  either been averaged over a Gaussian of width $R=2/Q_s(Y)$ centered
  at a random impact parameter or over the entire 2d impact parameter
  plane of a large lattice ($L\simeq18/Q_s(Y=0)$). The transverse area
  is taken as $A_\perp=2\pi R^2$ or $A_\perp=L^2$, respectively. A bar
  refers to an average over all gluons with transverse momentum
  $q>Q_s(Y)$ for a given configuration. Left: evolution rapidity $Y=0$
  corresponds to the MV model. Right: the fields have been evolved to
  $Y=1/\alpha_s$ via the JIMWLK equations.}
\label{fig:Ng-q2-scatter}
\end{figure}
In Fig.~\ref{fig:Ng-q2-scatter} we show the correlation of
$\Delta\overline{q^2}$ and $\Delta N_g$.  The MC data shows a tight
positive correlation of the transverse momentum vs.\ gluon density
fluctuation, as expected. The magnitude of the fluctuations of both
the gluon density per unit transverse area $\Delta N_g/A_\perp$ as
well as of the typical squared transverse momentum
$\Delta\overline{q^2}$ has increased by essentially an order of
magnitude from $Y=0$ to $Y=1/\alpha_s$; this is despite the fact that
both axes in Fig.~\ref{fig:Ng-q2-scatter} have been scaled by
$1/Q_s^2(Y)$ to make them dimensionless. It is also interesting to see
that in the MV model the fluctuations essentially scale with
``volume'', i.e.\ $\Delta N_g$ is approximately proportional to
$A_\perp$ while $\Delta\overline{q^2}$ is independent of $A_\perp$. At
$Y=1/\alpha_s$ on the other hand, the solution of JIMWLK clearly
exhibits finite range correlations since averaging over a large
``volume'' strongly reduces $\Delta\overline{q^2}$ at fixed density
$\Delta N_g/A_\perp$.

It is interesting to obtain a rough idea of the magnitude of $\Delta
N_g$ for reasonable values of $A_\perp$ and $Q_s(Y)$. A semi-hard
process may effectively average the target gluon fields over an area
of order $A_\perp\simeq0.1$~fm$^2$. Choosing a target saturation
momentum of $Q_s(Y)=1$~GeV we can then translate $\Delta N_g/(A_\perp
Q_s^2)=1$, 2, 3 on the horizontal axis of Fig.~\ref{fig:Ng-q2-scatter}
to $\Delta N_g\simeq5$, 10, 15 additional semi-hard gluons; for
$Q_s(Y)\simeq2.5$~GeV this increases to about $\Delta N_g\simeq30$,
60, 90 excess gluons in the target.

\begin{figure}[htb]
\begin{center}
\includegraphics[width=0.49\textwidth]{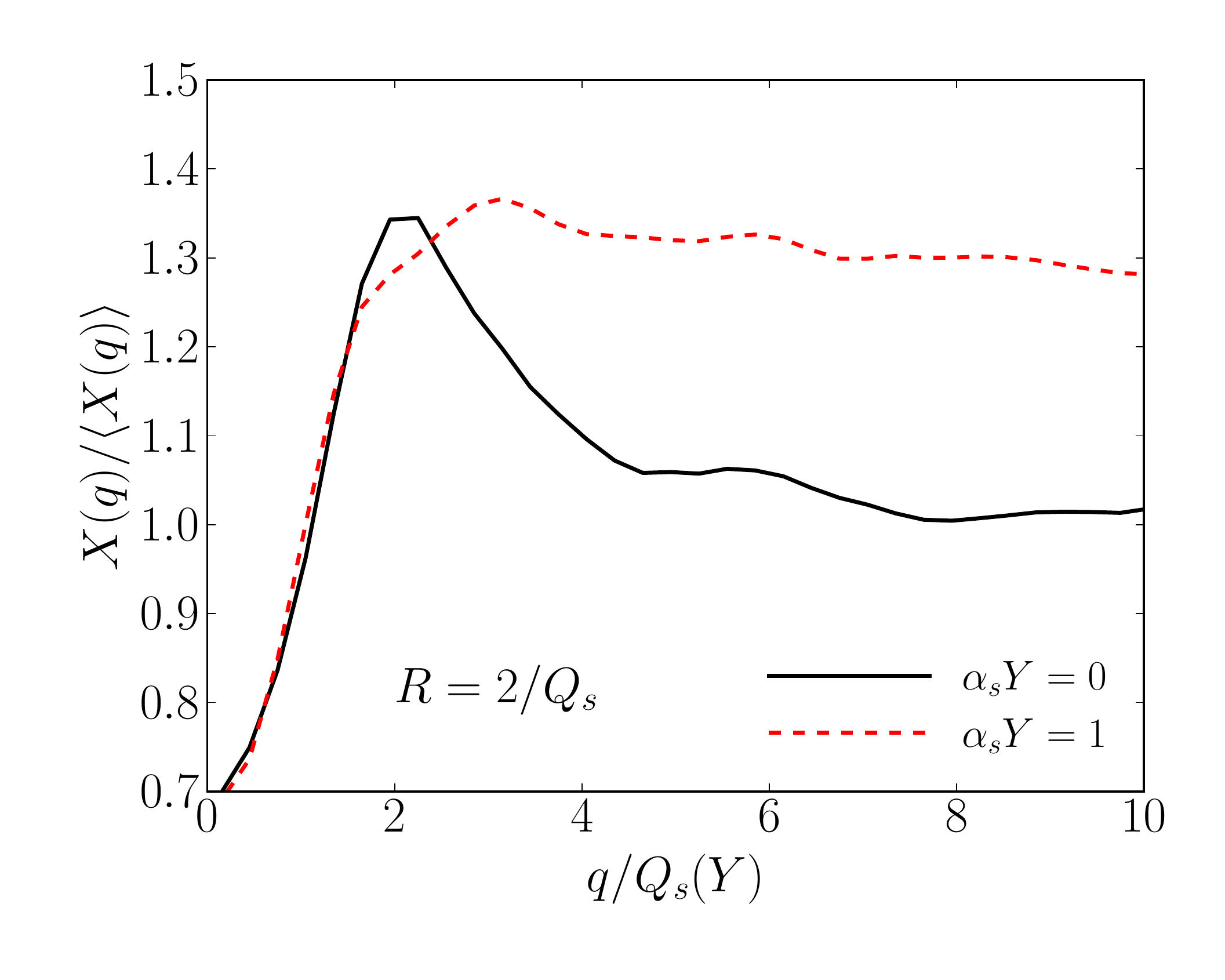}
\includegraphics[width=0.49\textwidth]{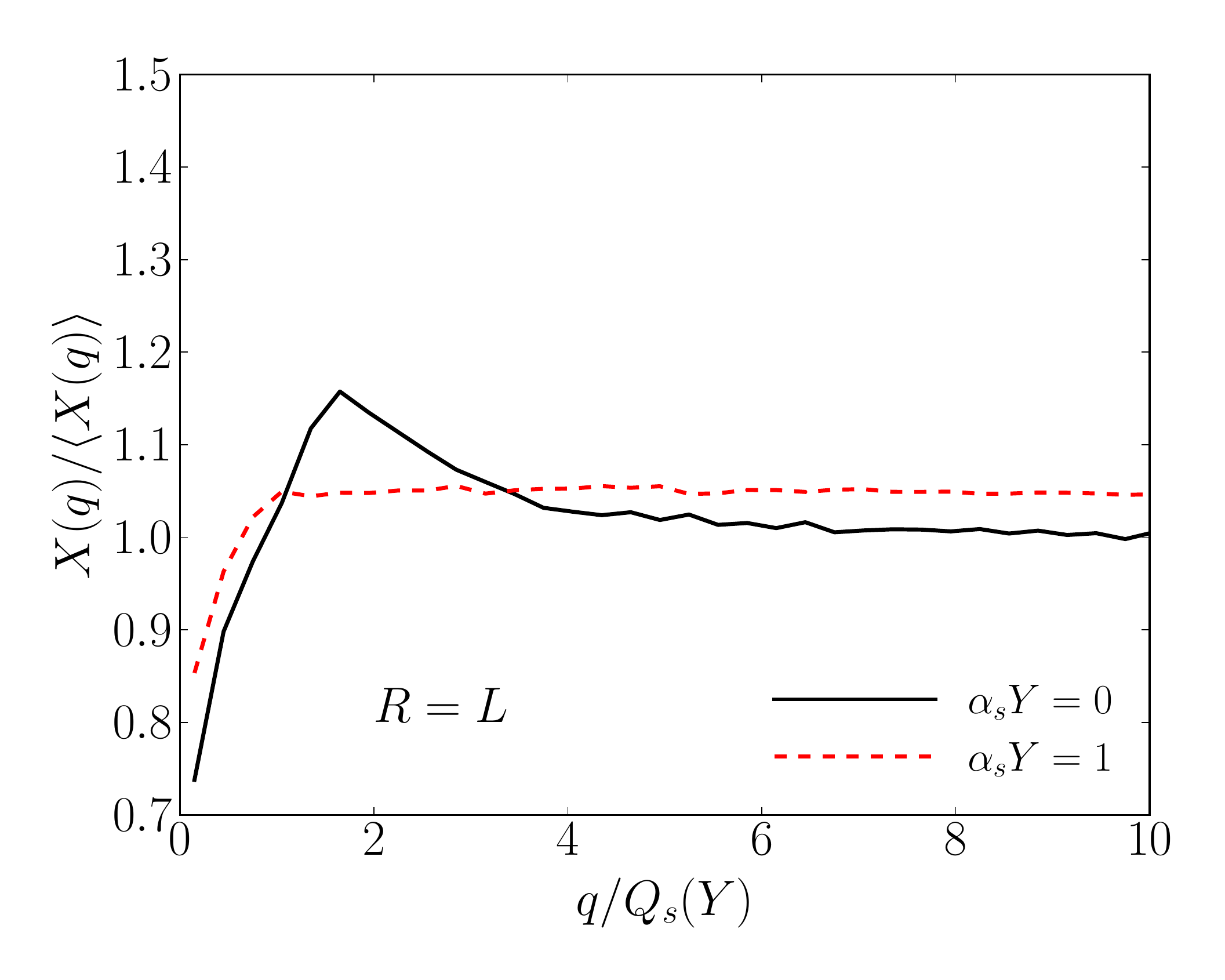}
\end{center}
\vspace*{-1cm}
\caption[a]{Fluctuation of the gluon distribution $X(q)=g^2 \tr
  |A^+(q)|^2$ in the MV model ($Y=0$, solid line) and after leading
  order, fixed coupling JIMWLK evolution to $\alpha_sY=1$ (dashed
  line). The left panel corresponds to the gluon distribution
  integrated over a Gaussian of width $R=2/Q_s(Y)$ centered at a
  random impact parameter; for the panel on the right the gluon
  distribution has been integrated over the entire 2d impact parameter
  plane of a large lattice ($L\simeq18/Q_s(Y=0)$). To obtain smooth
  curves we have averaged over a subsample of 100 configurations (out
  of 1000 total) with the highest gluon multiplicity $N_g$.  }
\label{fig:X_shape}
\end{figure}
The spectral shape of high-multiplicity fluctuations is shown in
Fig.~\ref{fig:X_shape}. For the MV model the dominant fluctuations
contain additional gluons with transverse momenta up to a few times
$Q_s$, then drop off smoothly to unity for $q\gg Q_s$. Qualitatively,
this tail corresponds to our {\sl ansatz}~(\ref{eq:eta(q)}) with $a>0$. At
high rapidity quantum fluctuations change the shape of fluctuations to
a flat, essentially scale independent distribution, so $a\approx0$ in
Eq.~(\ref{eq:eta(q)}). The different $q$-dependence of the
fluctuations illustrates the different role of the saturation scale
$Q_s$ in the MV model vs.\ JIMWLK evolution: in the MV model this
scale truly affects the dynamics of fluctuations which ``pile up''
just above $Q_s$. If the hadronic wave function evolves to much smaller $x$,
on the other hand, $Q_s$ is not a prominent scale in the fluctuation
spectrum but acts merely as an absorptive boundary for BFKL
emissions~\cite{Mueller:2002zm}. Indeed, recall that the canonical
dimension of $\eta(q)$ is zero and that the JIMWLK evolution
kernel at fixed coupling is scale invariant.

Fig.~\ref{fig:X_shape} also shows that as expected fluctuations in a
smaller ``volume'' have greater amplitude. Other than that the
spectral shape of $\eta(q)$ averaged over small ($R=2/Q_s(Y)$) or
large scales in the impact parameter plane is similar.

Amusingly, Fig.~\ref{fig:X_shape} resembles qualitatively the
``disappearance of the Cronin peak'' due to small-$x$
evolution~\cite{Albacete:2003iq}. Of course, the latter refers to the
{\em averaged} evolution of the {\em ratio} of the gluon distributions
of a dense to a dilute target. In contrast, Fig.~\ref{fig:X_shape}
shows the transverse momentum spectrum of fluctuations of the gluon
distribution of a single target about the average/extremal function.

\begin{figure}[htb]
\begin{center}
\includegraphics[width=0.49\textwidth]{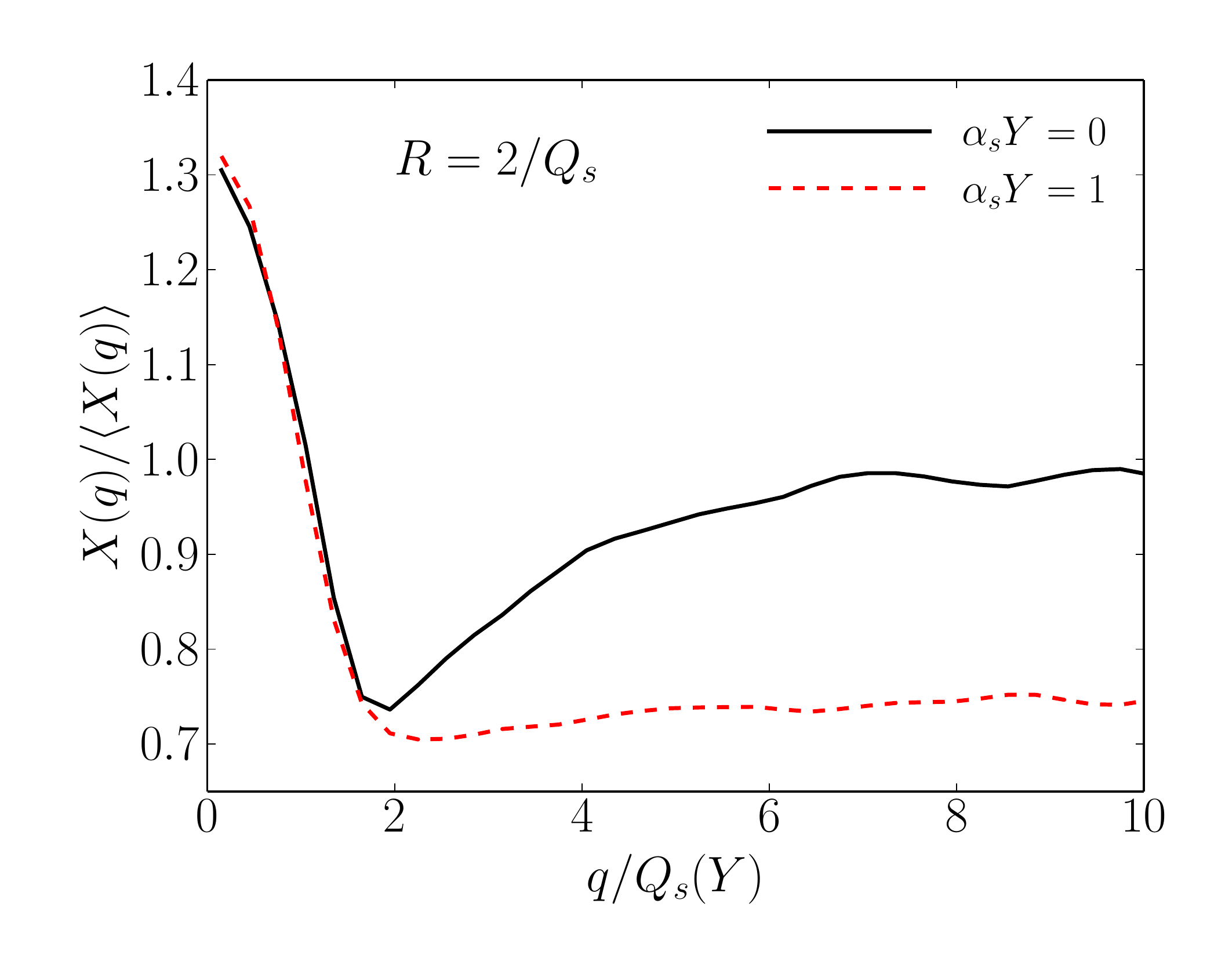}
\includegraphics[width=0.49\textwidth]{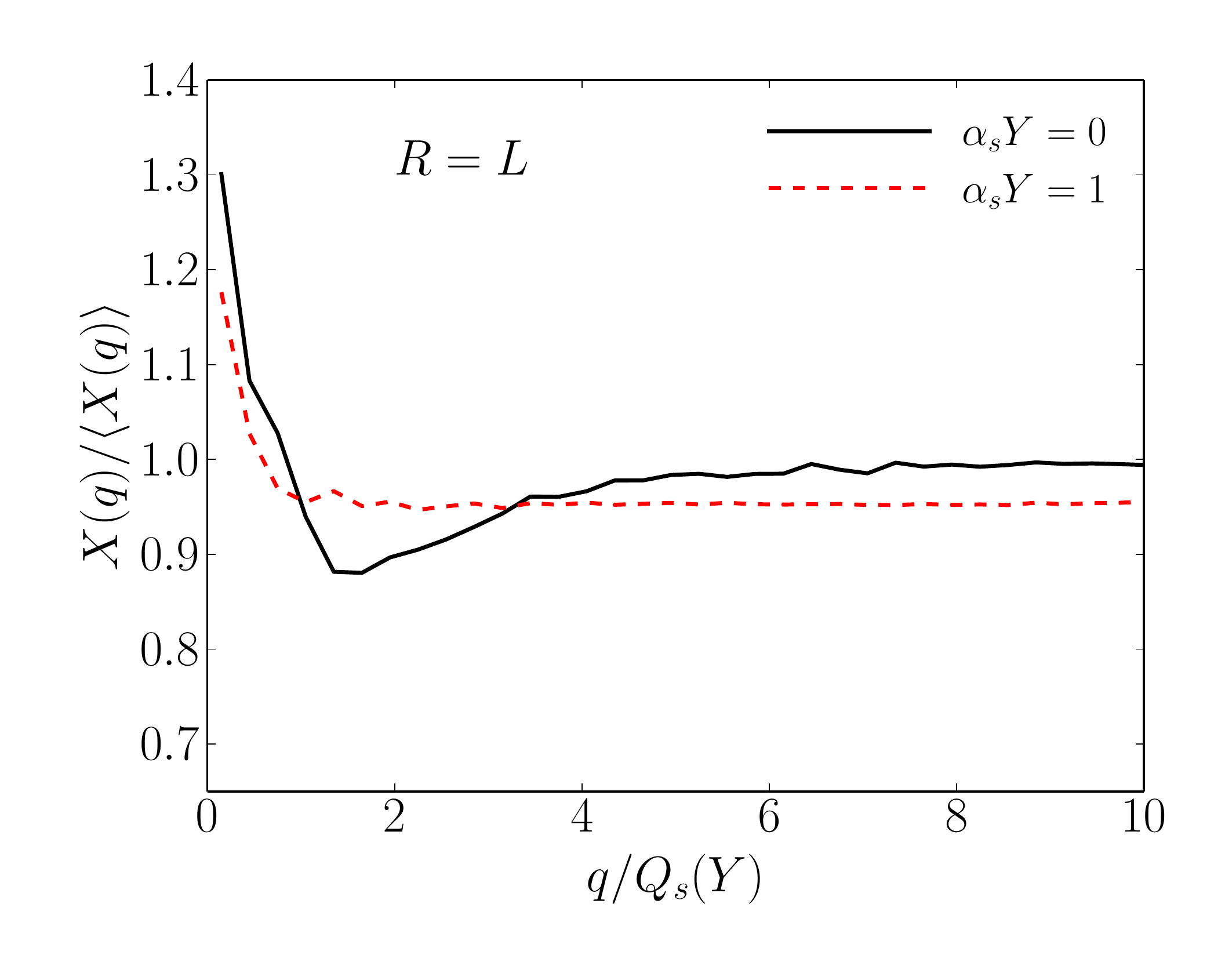}
\end{center}
\vspace*{-1cm}
\caption[a]{Spectral shape of low gluon multiplicity
  configurations. See Fig.~\ref{fig:X_shape} for further details.}
\label{fig:X_shape_low}
\end{figure}
Configurations with {\em lower} than average gluon multiplicity
exhibit fluctuations with a similar spectral shape as high
multiplicity configurations as shown in Fig.~\ref{fig:X_shape_low}. In
the MV model there is a dip in the gluon distribution just above the
saturation momentum, and the gluon distribution then smoothly
approaches the average distribution at higher $q$. On the other hand,
JIMWLK evolution again generates a scale invariant fluctuation and a
uniform depletion of gluons for transverse momenta greater than (one
or two times) $Q_s(Y)$.

\begin{figure}[htb]
\begin{center}
\includegraphics[width=0.49\textwidth]{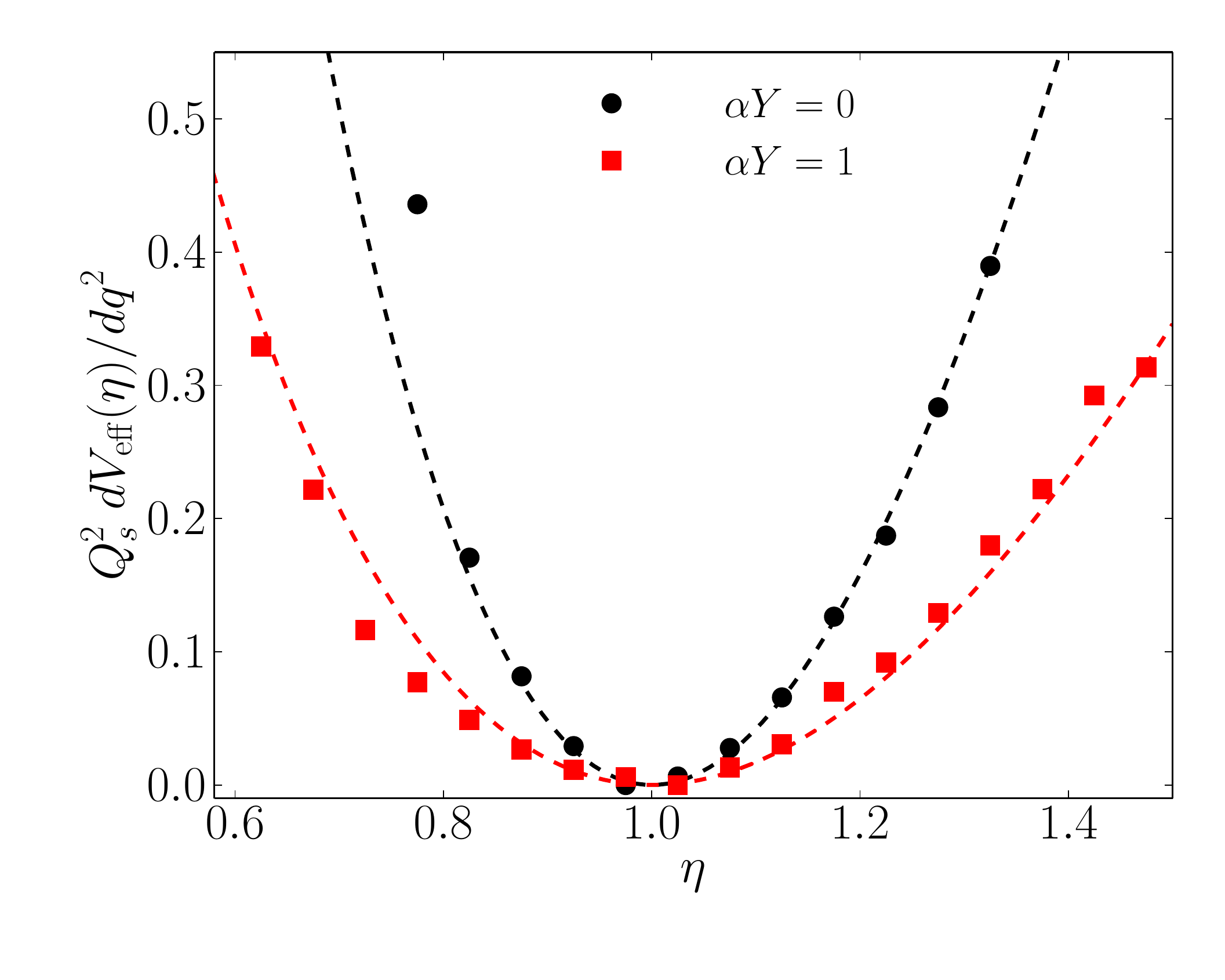}
\end{center}
\vspace*{-1cm}
\caption[a]{The effective potential describing fluctuations of the
  covariant gauge gluon distribution (beyond the saturation scale) in
  a transverse area patch of order $2\pi R^2 = 8\pi/Q_s^2(Y)$. Symbols
  show the results obtained from the MC simulation, lines correspond
  to the potential derived analytically (see text).}
\label{fig:Veff}
\end{figure}
We have also checked that the fluctuations of the gluon distribution
are indeed described by the Liouville potential (up to a field
redefinition) derived above in Eq.~(\ref{eq:Veff_eta}). In the MC
simulation this can be achieved by recording a histogram of
$\eta(q)=X_R(q,b)/\langle X_R(q,b)\rangle$ in the vicinity of an
arbitrary impact parameter. See appendix~\ref{app:VeffMC} for
details. We compare the simulation results to
\be \label{eq:Veff_MC_compare}
\frac{\d V_\mathrm{eff}}{\d q^2} =
\frac{1}{8\pi} N_c^2 A_\perp \left[\eta -1 -\log \eta\right]~.
\ee
The gluon distribution $X_R(q,b)$ has been ``smeared out'' over a
Gaussian of area $2\pi R^2$ centered at impact parameter $b$ as
described above. Nevertheless, the quantity $A_\perp$
in Eq.~(\ref{eq:Veff_MC_compare}) is a {\em dynamical} scale corresponding
to the transverse area occupied by the fluctuations $\eta(q)$ of the
gluon distribution. In particular, $A_\perp$ may very well be less
than $2\pi R^2$ if fluctuations occur over shorter length scales.

In Fig.~\ref{fig:Veff} we compare the numerical results to
Eq.~(\ref{eq:Veff_MC_compare}); $A_\perp$ in that equation has been
treated as a free parameter adjusted to best fit the MC
data\footnote{The simulation is carried out for $N_c=3$ colors while
  Eq.~(\ref{eq:Veff_MC_compare}) applies in the large-$N_c$
  limit. Subleading corrections simply rescale $A_\perp$. With the
  prefactor from Eq.~(\ref{eq:Veff_MC_compare}) the best fits
  correspond to $A_\perp Q_s^2\simeq23.75$ at $Y=0$, which is close to
  the geometric area $2\pi R^2 Q_s^2 = 8\pi$; and $A_\perp
  Q_s^2=10.22$ at rapidity $\alpha_sY=1$.}. Most significantly we
observe that the linear minus logarithmic potential from
Eq.~(\ref{eq:Veff_MC_compare}) indeed does describe the simulation
results rather well (within statistical uncertainties).

\section{Discussion and Outlook}
\label{Sec:conc}

In this paper we have described a new attempt at understanding
fluctuations of a physical observable $O[A^+]$, for example of the
(covariant gauge) gluon distribution $O[A^+]=g^2\tr |A^+(q)|^2$,
induced by fluctuations of the classical small-$x$ color fields. Of
course, fluctuations of the multiplicity in small-$x$ evolution have
been studied before, see for example the recent
paper~\cite{Liou:2016mfr} and references therein. These are typically
formulated in terms of dipole splitting processes. Instead, our
approach here involves the small-$x$ effective action $S[A^+]$,
resp.\ the weight functional $W[A^+]$. We stress that here $A^+$
refers to the soft {\em classical} field generated by integrating out
hard partons~\cite{MV,jimwlk} and representing them by random valence
charge sources. The semi-classical treatment of fluctuations requires
that one selects fluctuations which are suppressed by a probability
$p$ which is independent of the thickness $\sim A^{1/3}$ of the target
nucleus. The resulting fluctuation of the two-point function of $A^+$
is then proportional to the thickness, just like the extremal gluon
distribution itself. That is, such fluctuations of the small-$x$ field
are induced by longitudinally coherent fluctuations of the valence
charges as described (at moderately small $x$) by the MV model, and as
re-summed by JIMWLK evolution. In contrast, the treament of
fluctuations corresponding to a suppression factor $p\sim \exp\left(-
1/A^{1/3}\right)$ require higher-order corrections in the
coupling~\cite{Kovchegov:1999ua}.

Our approach allows us to discuss fluctuations even in the absence of
strong small-$x$ evolution, e.g.\ in the McLerran-Venugopalan
model. Furthermore, it can be applied to observables which may be more
difficult to access in dipole splitting approaches.  For example, we
can define, and in principle compute, the functional distribution of
the Weizs\"acker-Williams gluon distribution:
\be \label{Summary:Veff_WW}
e^{-V_\mathrm{eff}[X(q)]} =
\int {\cal D}\rho(q) \, W[\rho(q)]\, \delta(X(q)-g^2\tr |A^i(q)|^2)~.
\ee
We have computed this potential analytically in the weak field limit
($A^i \sim 1$), and for a large number of colors
($N_c\gg1$). Nevertheless, it is feasible, in principle, to compute it
from Eq.~(\ref{Summary:Veff_WW}) even when $gA^i \sim 1$ and for any
$N_c$, perhaps numerically. At next to leading order in the field
strength, for example, we have in terms of the covariant gauge field
\bea
\delta^{ij} \, g^2\tr A^i(q) A^j(-q) &=& \frac{1}{2} q^2 g^2
                 A^{+a}(q)A^{+a}(-q) \nonumber\\
& & \hspace{-2cm}  - \frac{g^4}{8} f^{abe} f^{cde}
  \left(\delta^{lm}-\frac{q^lq^m}{q^2}\right)
  \int \dd k \dd p k^l p^m A^{+a}(q-k)A^{+b}(k)A^{+c}(-q-p)A^{+d}(p)~,
  \label{Summary:xG}\\
\left(2\frac{q^iq^j}{q^2}-\delta^{ij}\right) 
\, g^2\tr A^i(q) A^j(-q) &=& \frac{1}{2} q^2 g^2 A^{+a}(q)A^{+a}(-q)\nonumber\\
& & \hspace{-2cm}  + \frac{g^4}{8} f^{abe} f^{cde}
  \left(\delta^{lm}-\frac{q^lq^m}{q^2}\right)
  \int \dd k \dd p k^l p^m
  A^{+a}(q-k)A^{+b}(k)A^{+c}(-q-p)A^{+d}(p)~. 
    \label{Summary:xh}
\eea
The first line is the conventional Weizs\"acker-Williams gluon
distribution, the second line is the so-called distribution of
linearly polarized gluons\footnote{For an introduction into these
  gluon distributions see, for example,
  Ref.~\cite{Boer:2016bfj}. Their expectation values, i.e.\ their
  values at the extremum of $V_\mathrm{eff}$, have been computed to
  all orders in $gA^+$ within the MV model~\cite{Metz:2011wb} as well
  as at small $x$~\cite{Dumitru:2015gaa}. Expectation values of other
  such ``transverse momentum dependent'' (TMD) gluon distributions at
  small $x$ have been computed in Ref.~\cite{Marquet:2016cgx}.}. The
fluctuations of these distributions can be determined by substituting
the r.h.s.\ of Eqs.~(\ref{Summary:xG},\ref{Summary:xh}) into the
delta-functional in Eq.~(\ref{Summary:Veff_WW}). An explicit analytic
calculation at next to leading order in $gA^+$ is complicated by the
fact that the corrections are non-local in transverse momentum
space.  We leave this computation for future work.

As an application of interest to us we have used our approach to
determine the fluctuations of the (covariant gauge) gluon distribution
$g^2\tr |A^+(q)|^2$. This allowed us to study the correlation of the
fluctuations of the number of gluons (above the saturation scale) and
of their typical transverse momentum squared\footnote{We stress that
  we consider the number or transverse momentum of gluons in a single
  hadron or nucleus and not multiplicity or transverse momentum
  fluctuations in a collision of two hadrons or nuclei. The latter has
  been investigated, for example, in
  Refs.~\cite{Schenke:2013dpa}.}. We find that these quantities are
very tightly correlated so that an increase (decrease) in the gluon
density per unit transverse area corresponds to an upward (downward)
fluctuation of the squared transverse momentum. The solution of the
JIMWLK small-$x$ RG exhibits a much stronger increase of $\Delta
\overline{q^2}$ with the gluon density $\Delta N_g/A_\perp$ in small
``volumes'' (transverse patches of size a few times $1/Q_s^2(Y)$),
presumably due to the presence of finite range correlations in the
impact parameter plane.

The shape of such high-multiplicity fluctuations in transverse
momentum space is modified significantly by JIMWLK evolution to small
$x$ as compared to the MV model. The latter adds hard gluons mainly
right above the saturation scale. On the other hand, the solution of
the small-$x$ renormalization group (in the JIMWLK approximation)
gives approximately scale independent multiplicative fluctuations. In
other words, the fluctuations that emerge in the small-$x$ limit are
better characterized as scale invariant fluctuations of a
dimensionless field which multiplies the average gluon distribution,
rather than as fluctuations of the absorptive
boundary~\cite{Mueller:2002zm} set by the saturation momentum. A
nearly scale invariant spectral distribution of high (or low)
multiplicity fluctuations are a clear signature for perturbative
quantum evolution with a conformal evolution kernel. 
It would be interesting to check the modifications of the
fluctuation spectrum in the running coupling JIMWLK or
in the full next-to-leading-log JIMWLK.

\appendix 
\section{Obtaining the effective potential from numerical MC
  simulations}
\label{app:VeffMC}
In this appendix we present more details on how the effective
potential presented in Fig.~\ref{fig:Veff} has been extracted
from the numerical simulation.

At a given rapidity, $Y$, for a given configuration of Wilson lines
$V_Y(x_\perp)$, we compute the observable $X_R(q,b=0)$ using
Eq.~\eqref{eq:X_R_q_b} on a square $N\times N$ lattice in
$q-$space. Next, we split $X_R(q)$ into bins of $q^2$ defined by
\begin{equation}
	q^2  = \frac{4}{a^2} \sum_{n=1,2} \sin^2 \frac{\pi i_n}{N}~,
\end{equation}
where $i_n$ denotes the lattice site in the $n$-direction and $a$ is
the lattice spacing.  We then compute the ratio $\eta(q^2) =
X_R(q^2)/\langle X_R(q^2) \rangle$ for each configuration in each
momentum bin. In each bin of $q^2$ we again construct a histogram of
the distribution of values of $\eta(q^2)$ as it fluctuates
configuration by configuration. This results in a two-dimensional
histogram of the number of counts $C$ as a function of $\eta$ and
$q^2$. The logarithm of the number of counts, modulo an additive
constant $\log{\cal N}$, is the differential effective potential with
negative sign, i.e.\
\begin{equation}
	\frac{\d V_{\rm eff}}{\d q^2}  = - \log \left( {\cal N}
        C(\eta,q^2) \right) ~. 
\end{equation}
The constant ${\cal N}$ is chosen such that $\frac{\d V_{\rm eff}}{\d
  q^2} =0$ at $\eta =1$.  We have checked that within numerical
uncertainties this shift of the potential is about the same in each
momentum bin within the range $4 <q^2/Q_s^2 < 50$. We have also found that in
this range $\frac{\d V_{\rm eff}}{\d q^2}$ is momentum independent,
within statistical uncertainties. This enabled us to
average over all momenta in this range. The resulting potential is
presented in Fig.~\ref{fig:Veff}.

\section*{Acknowledgements}
We thank Yu.~Kovchegov, A.~Kovner, A.~Mueller, and R. Venugopalan for useful comments at the RBRC 
Workshop ``Saturation: Recent Developments, New Ideas and Measurements''; April 26-28, 2017, Brookhaven National Laboratory.

A.D.\ gratefully acknowledges support by the DOE Office of Nuclear
Physics through Grant No.\ DE-FG02-09ER41620; and from The City
University of New York through the PSC-CUNY Research grant 69362-0047.

\end{document}